\documentclass[prb,twocolumn,showpacs,floatfix]{revtex4}
\usepackage{amsmath,amsfonts,amsthm,graphicx}
\usepackage{graphicx}
\usepackage{graphics}
\usepackage{dcolumn}
\usepackage{bm}

\newcommand{\beq}{\begin{equation}}
\newcommand{\eeq}{\end{equation}}
\newcommand{\beqnar}{\begin{eqnarray}}
\newcommand{\eeqnar}{\end{eqnarray}}
\newcommand{\bfig}{\begin{figure}}
\newcommand{\efig}{\end{figure}}

\begin{document}
\title{Nonadiabatic pure spin pumping in zigzag graphene nanoribbons with proximity induced ferromagnetism}

\author{Hosein Cheraghchi}
\email{cheraghchi@du.ac.ir} \affiliation{School of Physics,
Damghan University, 36716-41167, Damghan, Iran}
\date{\today}
\newbox\absbox

\begin{abstract}
By combining Floquet theory with Green's function formalism, we
present non-adiabatic quantum spin and charge pumping through a
zigzag ferromagnetic graphene nanoribbon including a
double-barriers structure driven weakly by two local $ac$ gate
voltages operating with a phase-lag. Over a wide range of Fermi
energies, interesting quantum pumping such as i) pure spin pumping
with zero net charge pumping, ii) pure charge pumping and iii)
fully spin polarized pumping can be achieved by tuning and
manipulating driving frequency in the non-adiabatic regime. Spin
polarized pumping which is measurable using the current technology
depends on the competition between the energy level spacing and
driving frequency.
\end{abstract}


\maketitle
\section{Introduction}
Time-periodic gate voltages applied on a conductor can generate a
$dc$ pumped current even at zero bias voltage
\cite{Brouwer,Switkes,kohler,Marcus}. Quantum charge pumping can be
produced when the coherence length of electrons in mesoscopic
conductors (e.g. graphene and nanotubes) is larger than the device
length. On the other hand, additional to several micron mean-free
path of electrons, graphene has also very weak spin-orbit coupling
\cite{spinorbit} and, its spin flip length is so long about $1\mu
m$ at room temperature\cite{tombros}. So graphene-based devices
have a good opportunity for spintronic applications\cite{cheraghchi2}. However, weak
spin-orbit in graphene results in a weak electrical control over
spin polarization. With such large spin and phase coherence length
of electrons in graphene, it is expected that time-periodic gate
voltages can generate significant spin pumped currents passing
through ferromagnetic graphene\cite{zhang}. Recently, by improving
experimental techniques, spin splitting of electrons in graphene
is induced by strong proximity of a ferromagnetic insulator layer
which is deposited on graphene sheet\cite{Haugen} such that its
large mobility of carriers is preserved\cite{PRL2015}. Here we
apply a time-periodic electrostatic potential to control spin and
charge transport.
\begin{figure}
\centering
\includegraphics[width=9cm]{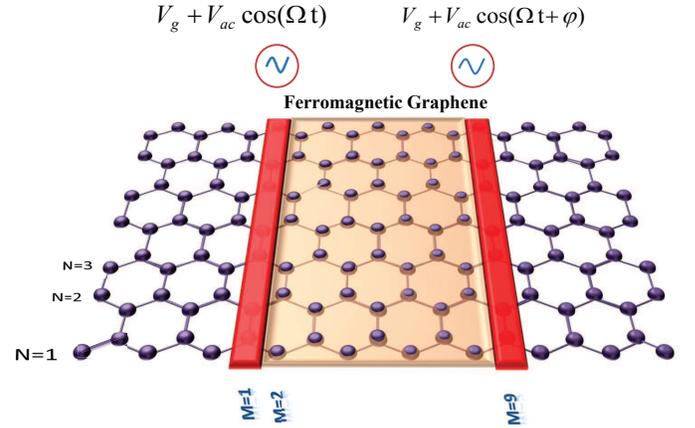}
\caption{ Schematic view of zigzag graphene nanoribbon ZGNR(M,N) with $M$ and $N$ atomic layers in
length and width, respectively. The reservoirs are considered to
be the same as the sample located at the central region.
Nanoribbons are driven by two periodic-time dependent gate
voltages with amplitude $V_{ac}$, frequency $\Omega$ and phase
difference $\varphi$. Two large barriers ($V_g$) are applied in
the first and last unit cells. }\label{schematic-view}
\end{figure}

The peculiar properties of bulk graphene such as
chirality\cite{RMP} induce surprisingly large pumped current
originating from evanescent modes at the Dirac point which is
larger than the pumped current measured in
semiconductors\cite{schomerus,pump-BG}. On the other hand,
controlling of Van-Hove singularities, ribbon size, the edge
structure and also defects make graphene nanoribbons (GNR) and
also nanotubes as the appropriate candidates for enhanced pumped
current
\cite{Foa-APL2011,JPCM,pump-GNR,Arrachea2012,GNR-Arm,Foa-PRB}.
Furthermore, spin current in monolayer graphene has been also
proposed to generate by application of two oscillating gate
voltages through the adiabatic quantum pumping in some special
Fermi energies\cite{zhang}. However, the main point of the present
work is the control of spin current by tuning of the driving
frequency in non-adiabatic regime. Moreover, in this regime, it is
possible to tune pumped current by using the other electrical
parameters
\cite{Moskalets,wang,Arrachea,Mahmoodian,Faizabadi,Moskalets2008}.
In this regime, the possibility of inducing Floquet topological
properties by laser illumination has recently opened a new
interesting field\cite{kitagawa}.

In this paper, by using Floquet Landauer
formalism\cite{kohler,Foa}, we investigate a crossover from
adiabatic to non-adiabatic charge and spin pumping in
ferromagnetic graphene nanoribbons induced by two weakly local
time-periodic gate voltages operating with a phase difference.
Depending on the system parameters, it is shown that a pure spin
and charge pumping emerges at some special driving frequencies. At
low frequency regime, spin splitting is not remarkable such that
there is no pure spin pumping at all, while at high frequency
regime and also in the presence of the resonant states arising
from the barriers, the situation for emerging pure spin pumping is
achievable. At the end, we present a discussion about practical
implementations of a quantum pump based on graphene nanoribbons.

This paper is organized as the following: In section II, we
present the methodology for calculating pumped current by using
Floquet approach in combination with Green's function formalism
(Floquet Landauer formalism). Then in section III, we present our
results on charge and spin pumping for zigzag graphene
nanoribbons. Finally, the last section includes the conclusions.
\section{Formalism}
As shown in Fig.\ref{schematic-view}, we consider zigzag graphene
nanoribbon in the proximity of ferromagnetic insulator such as
$EuO$ \cite{Haugen} and yttrium iron garnet (YIG)\cite{PRL2015}
which is connected to two infinite electrodes. For simplicity, the
electrodes are considered to be the same nanoribbon as the central
system. One $\pi$ orbital is considered per each site for graphene
system. So the single electron Hamiltonian of the central region
is described as

\beq\begin{array}{r} H^{\sigma}_{C}(t)=\sum_{i\sigma}
(\varepsilon^{C}_i(t)-\sigma h) c^{\dagger}_{i\sigma}c_{i\sigma}+
\\ \sum_{<ij>\sigma}t_0
(c^{\dagger}_{i\sigma}c_{j\sigma}+c_{i\sigma}c^{\dagger}_{j\sigma})
 \end{array} \label{hamiltonian}\eeq.

where $c^{\dagger}_{i\sigma}$ and $c_{i\sigma}$ are the electron
creation and annihilation operators and $t_0$ is the hopping
energy between nearest neighbour atoms, respectively. All energies
are scaled in unit $t_0$. $\sigma=\pm1$ refers to up and
down-spin. Here $h$ is the induced exchange field arising from
strong proximity of a ferromagnetic insulator. This induced
exchange field can be tuned by means of an in-plane external
electric field \cite{ex-tune}. The exchange splitting inducing by
a ferromagnetic insulator such as $Euo$ in graphene has been
estimated to be in the range of $5-65
meV$\cite{Haugen,semenov,PRL2015}. The coupling Hamiltonian which
is independent of time is defined as the following:
$H_{coupling}=\sum_{<ij>}(t_{LC}
c^{\dagger}_{i}c_{j}+t_{RC}c_{i}c^{\dagger}_{j})+H.c.$, where
$t_{LC}$ and $t_{RC}$ are the coupling energy between the
electrodes and central ribbon. To have quantum pumping, one should
break left-right symmetry by application of a dynamic or static
asymmetric spatial factor. Here, we will apply a dynamical break
of the left-right symmetry appearing as a phase lag between two
oscillating gate voltages. Therefore, in the case of static
symmetry such as $t_{LC}=t_{RC}$, one can still observe $dc$
pumped current. Harmonically time-dependent gate voltages are
applied just on the last and first atomic layers of the central
region as shown in Fig.\ref{schematic-view}.

\beq\begin{array}{r} \varepsilon^{C}_i(t)=eV^g_i+eV^{ac}_i
cos(\Omega t+\varphi_i)
 \end{array}\eeq.

where $V^{ac}_i$ and $\Omega$ are the strength and frequency of
the oscillating gate voltages, respectively. Moreover, let us
apply zero gate voltage on all sites $V^g_i=0$, except the first
and last unit cells that include two barriers as strong as
$V^g_i=1$. Constant gate voltage causes to shift Van Hove
singularities into the band center giving rise an enhanced pump
current\cite{Foa-APL2011}. To break the left-right symmetry, as
shown in Fig.\ref{schematic-view}, two $ac$ gate voltages with a
phase lag $\varphi=\varphi_R-\varphi_L$ are applied on the first
($\varphi_L=0$) and last ($\varphi_R=\varphi$) unit cells along
the transport direction. We consider the phase difference between
two $ac$ gate voltages to be $\varphi=\pi/2$ in which the maximum
$dc$ current has been reported in quasi-one dimensional
systems\cite{Moskalets,Arrachea,Arrachea2012}. Two semi-infinite
electrodes are not driven by any time-dependent gate voltages.
Because of zero source-drain voltage, we set on-site energies of
the electrodes to be zero.

To obtain $dc$ current through graphene nanoribbon, we use a
Floquet theory in combination with non-equilibrium Green's
function formalism. It should be noted that the electron-electron
interaction is not considered in this calculation. At zero
temperature, the averaged pump current over one period of the $ac$
gate voltage for spin $\sigma$ is written as the
following\cite{kohler,kitagawa,Foa}.

\beq\begin{array}{r}\overline{I}_{\sigma}=\frac{e}{h}\sum_{n}
\int_{-\infty}^{E_F} d\varepsilon T^n_{\sigma}(\varepsilon)
\end{array} \label{current}\eeq
where
$$T^n_{\sigma}(\varepsilon)=T_{RL,\sigma}^n(\varepsilon)-T_{LR,\sigma}^n(\varepsilon)$$
and $E_F$ is the Fermi energy. The transmission probability of
carriers coming from the left to the right electrode
$T_{RL,\sigma}^n(\varepsilon)$ assisted by the absorption ($n>0$)
or emission ($n<0$) of $\mid n \mid$ photons with an incident energy
of carriers $\varepsilon$ and spin $\sigma$ can be written in
terms of Floquet Green's function $G_{\sigma}^{n}(\varepsilon)$
as the following.

\beq
\begin{array}{r}T_{RL\sigma}^n(\varepsilon)=Tr[\Gamma^{\sigma}_{R,n}(\varepsilon) G_{RL,\sigma}(\varepsilon)
\Gamma^{\sigma}_{L,0}(\varepsilon)G_{RL,\sigma}^{\dagger}(\varepsilon)]
\end{array}
\eeq

where Floquet Green's fucntion is calculated in based on the
Floquet states as $G_{RL,\sigma}^{n}(\varepsilon)=\langle R,n
|(\varepsilon I-H_F)^{-1}|L,0\rangle$ and Floquet Hamiltonian is
defined as $ H_F = H_C(t)-i \hbar \frac{\partial}{\partial t}$.
The Floquet Green's function is a Fourier transformation of
retarded Green's function as the following:
$$G_{RL,\sigma}^r(t,\varepsilon)=\sum_{n=-\infty}^{\infty} G_{RL,\sigma}^{n}(\varepsilon)\exp(-i n \Omega t).$$

Taking Fourier transform of the equation of motion governing on
the retarded Green's function results in an equation of motion
followed by the Floquet Green's function. The escape rates of
electrons to the electrodes through elastic or inelastic channels
are defined as the following.

$$\Gamma_{\alpha,n}(\varepsilon)=i[\Sigma_{\alpha}(\varepsilon+n\hbar\omega)-\Sigma^{\dag}_{\alpha}(\varepsilon+n\hbar\omega)]$$

where the index $n$ shows the number of photons assisted for
inelastic transport and $\Sigma_{\alpha}(\varepsilon)$ is the
Floquet self energy arising from the electrodes $\alpha=L,R$. In
the calculations, the convergence of the transmission functions
determines the number of required Floquet channels $n$. In the
case of non-interacting systems, the above formalism is
equivalent to non-equilibrium Green's function formalism (Keldysh
formalism)\cite{Arrachea-moskalet}.
\begin{figure}
\centering
\includegraphics[width=8cm]{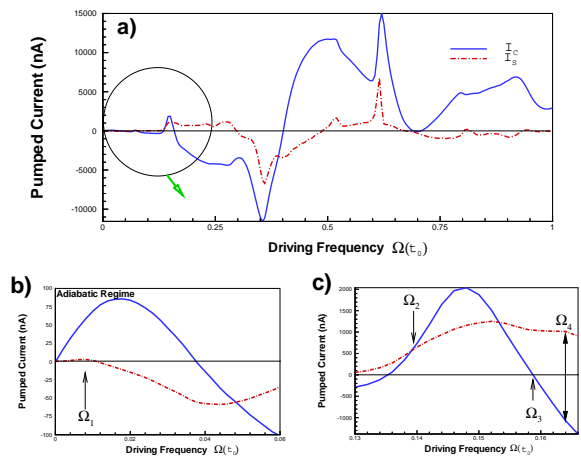}
\caption{Pumped current in terms of driving frequency at the band
center $E=0$ for ZGNR(12,4). Blue solid line (red dash-dot line)
shows charge (spin) pumped current. In the lower panels, some of
frequency's ranges are magnified to present some special
frequencies.}\label{Zomega-even}
\end{figure}

\section{Results}
\subsection{Driving Frequency Dependence of Pumped Current}
We present in Fig.\ref{Zomega-even} the results for spin and
charge pumped current through zigzag graphene nanoribbons as a
function of driving frequency $\Omega$ at the band center. The
charge and spin pumped current are defined as
$I_{C}=I_{\uparrow}+I_{\downarrow}$ and
$I_{S}=I_{\uparrow}-I_{\downarrow}$, respectively. $I_{\uparrow}$
and $I_{\downarrow}$ refer to the up and down spin currents. For
very low frequencies (adiabatic regime), there is a linear
behavior of charge pumped current in terms of $\Omega$ which is in
compatible with the results reported before\cite{pump-GNR}. This
linear behavior which is shown in Fig.(\ref{Zomega-even}.b), is a
character of the adiabatic pumping\cite{Brouwer}. However, close
to the Van Hov singularities, the scale of pumped current with
the frequency is improved\cite{Foa-APL2011}.

The characteristic energy scale in this system is defined as
$n_{max}*\Omega$ in which $n_{max}$ is the maximum number of
sidebands excited by the driving gate voltages. The adiabatic
regime occurs when this energy scale is small enough in compared
to the level spacing of the resonant states. On the other hand,
$n_{max}$ depends on the ratio of $V_{ac}/\Omega$. It means that
the adiabatic regime occurs at lower frequencies when the
strength of pumping is large. Furthermore, longer or wider
ribbons result in smaller level spacing. Therefore, larger pumps
have smaller frequency ceiling for working in the adiabatic
regime. The adiabatic regime is indicated by the point $1$ in
Fig.(\ref{Zomega-even}.b) which corresponds to the pumping
frequency as small as $\Omega_1=0.01t_0$. Although the level
spacing is energy dependent, in our system, one can estimate it
to be about $0.2$ ($t_0$) which is larger than $\Omega=0.01
(t_0)$. In the adiabatic regime, spin pumping is small, however
there is no pure spin pumping in this regime.

At high frequency regime, spin and charge pumped current show some
extremum points at the special values of the driving frequencies
which are originated from the resonance of the Floquet states with
those atomic levels captured between two barriers. Therefore one
can control the direction and amplitude of the spin current by
tuning the driving frequency. At frequencies that are multiples of
the level spacing, by mediated of photon-assisted transitions,
atomic levels are mixed with each other. Therefore, an enhanced
pumped current emerges in this situation. However, level spacing
is not uniform along the whole spectrum, so an interplay between
driving frequency and level spacing is more complicated.

In the non-adiabatic regime, spin-polarized pumped current is
observable along the whole range of driving frequencies shown in
Fig.(\ref{Zomega-even}). However, full spin polarized pumping are
categorized in the following three interesting cases: i) in the
case of $I_{C}=I_{S}$, there exists pure up-spin polarized
current. So we have $I_{\uparrow}\neq 0,I_{\downarrow}=0$. This
case is marked by the point $2$ in Fig.(\ref{Zomega-even})
referring to $\Omega_2$. If charge and spin pumped current are
positive/negative, up-spin current is pumped from the left/right
to the right/left electrode. ii) The case of $I_{C}=0$ which
emerges at the frequency $\Omega_3$, results in a fully spin
pumping with zero net charge pumping. The pumped currents for up
and down spins are in the opposite directions
$I_{\uparrow}=-I_{\downarrow}$.

In fact, quantum interference is responsible for such interesting
cases. In the adiabatic regime, pumped current is an odd function
of the phase difference between oscillating gate voltages.
However, in the non-adiabatic regime, an additional contribution
to the phase difference is originating from the spatial phase
difference of the Floquet sidebands\cite{Moskalets}. This phase
difference which is absent at low frequency regime, is responsible
for the sign reversal of the pumped current at consecutive peaks
shown in Fig.\ref{Zomega-even}. Photon-assisted Quantum
interference through different Floquet side-bands can result in a
special case; two spin-polarized pumped current which flow in the
opposite directions. So in this case, there is no net charge
pumping while it is purely spin-polarized pumping in the opposite
directions.

iii) Correspondingly, the case of $I_{C}=-I_{S}$ also results in a
fully down-spin polarized currents $I_{\uparrow}=
0,I_{\downarrow}\neq 0$. If $I_C>0$, then we have
$I_{\downarrow}>0$. This frequency is indicated by $\Omega_4$. In
other cases, the pumped current is partially spin polarized. Spin
polarization emerges when the density of state for up and down
spins shift against each other, so at a given frequency, level
spacing for up spins differs from one for down spins. As a
consequence, pumped current for up and down spin would be
different in its value and also pumping direction. Moreover the
effect of higher orders of the orbital overlaps on the pumped
current is investigated in the Appendix.
\subsection{Fermi Energy Dependence of Pumped Current}

\begin{figure}
\centering
\includegraphics[width=9cm]{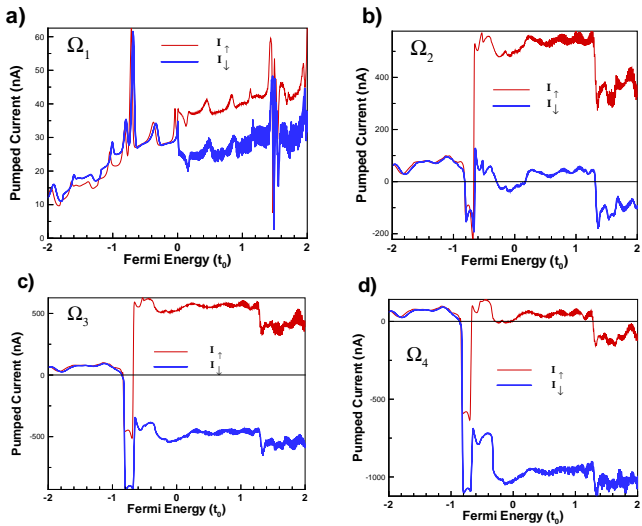}
\caption{ Pumped current in terms of Fermi energy for ZGNR(12,4)
at the specified driving frequencies of
$\Omega_1,\Omega_2,\Omega_3,\Omega_4$ which have been marked in
Fig.\ref{Zomega-even}.}\label{Zenergy-even}
\end{figure}

\begin{figure}
\centering
\includegraphics[width=8cm]{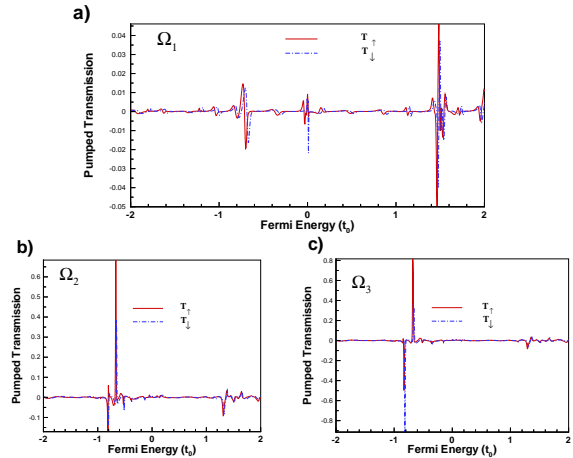}
\caption{ Net pumped transmission in terms of Fermi energy for
ZGNR(12,4) at the specified driving frequencies of
$\Omega_1,\Omega_2,\Omega_3$ which have been marked in
Fig.\ref{Zomega-even}. The norm of the left-right green's function
for single photon-assisted transmission is plotted in part (b) for
the low frequency regime $\Omega_1$. }\label{cond-zigzag}
\end{figure}

Let us concentrate on the Fermi energy dependence of the pumped
current at these four special driving frequencies shown in
Fig.\ref{Zenergy-even}. Panel (a) in Fig.\ref{Zenergy-even} shows
spin-up and down pumped current as a function of Fermi energy for
small pumping frequency $\Omega_1=0.01t_0$. This frequency is
smaller than two consecutive energy levels of the central region.
The pumped current contains successive small peaks which are
modulated on a pumped current\cite{Foa-APL2011}. These peaks in
the pumped current are originated from the peak-antipeak
structure\cite{Arrachea} of the net pumped transmission ($\sum_n
T^n_{\sigma}$) shown in Fig.(\ref{cond-zigzag}.a). In this regime,
there is no mixing between system levels, therefore spin
polarization is small. However, there exists unidirectional charge
pumping along the whole range of Fermi energies.

Panels (b), (c) and (d) in Fig.\ref{Zenergy-even} indicate fully
spin polarized pumped current in terms of Fermi energy for the
frequencies $\Omega_2,\Omega_3,\Omega_4$. Apparently, because of
two barriers applied on the first and last unit cells,
electron-hole symmetry breaks. Furthermore, the pumped current
increases around $E=-1$ which is in resonance with the barriers.
In panel (b) and at $\Omega_2$, down-spin current is blocked and
up-spin is fully pumped from the right to the left electrode. At
$\Omega_3$ and in the panel (c), we present an interesting case of
pure spin current with zero net charge pumping over a wide range
of Fermi energy around the band center. Finally, at the frequency
$\Omega_4$, down-spin is fully pumped, while up-spin is still
blocked. There is some structure of jumps and plateaus in panels
(b) to (d) which follow the resonant states shown in the net
pumped transmission curve (Fig.\ref{cond-zigzag}b,c). The behavior
of pumped current in ZGNRs is similar to the pumped current
passing through one-dimensional chain\cite{Arrachea}.

Fig.(\ref{cond-zigzag}) shows the net pumped transmission curve
(which is defined as $\sum_n T^{n}_\sigma(\varepsilon)$) as a
function of energy for different driving frequencies. For low
driving frequency $\Omega_1$, peaks and anti-peaks in the net
transmission curve emerge around the resonant states located
between two barriers (Fig.(\ref{cond-zigzag}.a)). At both energies
of peak and anti-peak, we observe large quantum pumping and hence
constructive interference. On each peak-antipeak pattern, at lower
energies, the left to the right pumping is dominant while at
higher energies, pumping direction will be from the right to the
left side. The following adiabatic picture may be useful to
understand the reason for such behavior: At half period of
oscillating potential, the left barrier has a lower gate voltage
in compared to the right barrier. So electrons flow from the right
reservoir into the well. In the other half of period, the right
barrier gets lower which leads to a current flow from the well
into the right reservoir. If we turn off the oscillating gates,
transmission would have peaks at the resonant states as the
original transporting channels.

At low frequency regime, there is no mixing between system levels,
so transport occurs mainly through resonant and also one Floquet
side state. However, at high frequencies, the system levels are
mixed with each other arising from the driven potential. As a
result, large peaks are appeared in the net pumped transmission
which are observed in Figs.(\ref{cond-zigzag}.b,c).

\subsection{Practical Implementations}

In reality and in the absence of coulomb blockade, stray
capacitances between the gates and electron reservoirs may affect
adiabatic quantum pumping in the work done by
Switkes\cite{Switkes}. However, some features of the observed dc
voltage are in agreement with the theoretical calculations based
on the scattering matrix\cite{Brouwer2001}. To have conclusive
spin pumping, one of the powerful techniques is precessing
magnetization at ambient temperature\cite{bauer-RMP}.
Furthermore, the other technique is the use of the ac Josephson
effect in a hybrid normal-superconducting system \cite{russo}or
charge pumping in an unbiased InAs nanowire embedded on a
superconductor\cite{nature-josephson}. For graphene nanoribbons
with smooth edges, in the adiabatic regime, it is expected to
observe some general feature of quantum pumps at low temperatures.
Furthermore, we have shown before\cite{cheraghchi} that in
graphene nanoribbons, under application of a weak source-drain
voltage, screening effects are weak. Therefore, in a weak driving
gate voltages, accumulation and depletion of charge would be
small. On the other hand, homogeneity of the junctions reduce
stray capacitances between oscillating gates and electron
reservoirs. Here we assume that the electrodes are similar to the
pumping region for having transparent contacts in compared to the
normal pumps.

In reality, the width of nanoribbons are much wider than what we
have considered in this paper. Wider and longer nanoribbons have
smaller energy level spacing. So it is expected that for wider and
longer nanoribbons, more sign reversal happens for the pumped
current curve in terms of the driving frequency. For graphene
sheet, there exists pure spin pumping at special energies in the
adiabatic regime\cite{zhang}. Pure spin pumping is also expected
to observe in wider graphene nanoribbons.

During fabrication processes, the ribbons are full of impurity or
disorder. So we propose to investigate the effect of disorder and
impurity locations on the pure spin and charge pumping through
graphene nanostructures.

\section{Conclusion}
In conclusion, we study charge and spin pumping in ferromagnetic
graphene nanoribbons including a double-barrier structure driven
weakly by two local {\it ac} gate voltages with a phase-lag. With
new improvements to induce exchange magnetic field in graphene
\cite{PRL2015}, the main point of this paper is that charge and
spin pumping can be well controlled by tuning the driving
frequency in the non-adiabatic regime. Furthermore, fully spin
pumping with favorite direction is achievable at the special
frequency of {\it ac} gate voltages. It seems that at low
frequency regime, single photon-assisted tunneling is dominant,
while at higher frequencies, mixing of energy levels arising from
the driving field leads to constructive interference.
\begin{figure}
\centering
\includegraphics[width=8cm]{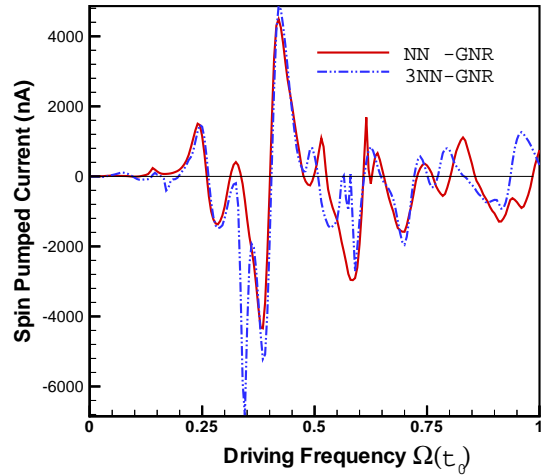}
\caption{ Spin pumped current ($I_S$) as a function of driving
frequency for ZGNR(20,4) at the band center for two cases: a) by
considering of the nearest neighbor overlaps (NN-GNR), b) the
second and third nearest neighbor hopping energies
(3NN-GNR).}\label{3NN-pump}
\end{figure}

\section{Acknowledgement}
We gratefully acknowledge Luis E. F. Foa Torres for his useful
comments during improvement of this work. HC thanks the
International Center for Theoretical Physics (ICTP) for their
hospitality and support during a visit in which part of this work
was done.

\section{Appendix}
\subsection{Higher order orbital overlaps}

In the tight binding model shown in equation \ref{hamiltonian},
hamiltonian contains nearest neighbor interaction. However, {\it
ab initio} band structures can be extracted by considering of the
second and third nearest neighbor overlaps in Hamiltonian. To
investigate the effect of higher order overlaps, the pumped
current is calculated by using tight-binding parameters which are
proposed by Ref.(\onlinecite{Riech}). The distances between the first,
second and third nearest neighbors are given by $a_1=1.42 {\hat{A}},
a_2=\sqrt{3} a_1, a_3=2a_1$. The values of nearest-neighbor
hopping parameters are calculated by comparing and fitting the
first principle and tight-binding band structures for different
types of graphene\cite{Roche} (2D graphene, graphene nanoribbons and nanotubes
). The second and third hopping parameters are
considered as $0.12 eV$ and $0.068 eV$ if the nearest-neighbor
hopping $t_0$ is considered to be as $2.78 eV$\cite{Riech}.

Fig.\ref{3NN-pump} shows spin pumped current ($I_S$) in terms of
driving frequency for the two cases: Considering of a)the nearest
neighbor overlap (NN-GNR), b) the second and third nearest
neighbor overlap (3NN-GNR). As it is shown, spin pumped current
shows more extremum for the 3NN-GNR case in compared to the NN-GNR
case.

\end{document}